\documentclass[10pt,aps,prd,superscriptaddress,twocolumn,amsmath,amssymb,floatfix,
nofootinbib 
]{revtex4-1} 

\usepackage[dvips]{graphics}
\usepackage{bm}
\usepackage{epsfig}
\usepackage{enumerate}
\usepackage{subfigure}
\usepackage{color} 
\usepackage{graphicx}
\usepackage[dvipsnames]{xcolor}
\usepackage{graphicx}
\usepackage[colorlinks,citecolor=blue,linkcolor=blue,urlcolor=blue,plainpages=false,pdfpagelabels]{hyperref}

\newcommand{\beq}{\begin{equation}}
\newcommand{\eeq}{\end{equation}}
\newcommand{\bea}{\begin{eqnarray}}
\newcommand{\eea}{\end{eqnarray}}

\begin{document} 
\title{Collective modes in imbalanced nodal line semimetal}

\author{SK Firoz Islam}
\affiliation{Department of Applied Physics, Aalto University, P.~O.~Box 15100, FI-00076 Aalto, Finland}

\author{Alexander A. Zyuzin}
\affiliation{Department of Applied Physics, Aalto University, P.~O.~Box 15100, FI-00076 AALTO, Finland}
\affiliation{Ioffe Physical--Technical Institute,~194021 St.~Petersburg, Russia}

\begin{abstract}
In this work, we investigate collective modes in a nodal line semimetal with two nodal-lines that have opposite spin polarization in the presence of spin population imbalance.
We find the components of polarization operator taking into account the electron-electron exchange interaction and obtain the dispersion relations of collective modes for the bulk and surface states. 
There exist four modes in the bulk, among which one is gapless and other three are gapped. The gapless surface mode is sensitive to the boundary conditions. 
\end{abstract}  
\maketitle
{\em Introduction}
Recently, three dimensional Dirac materials where conduction and valence bands merge along a line, so called nodal line (or ring) semimetals (NLSM),
have attracted intense research interests due to their unique band structure \cite{PhysRevLett_shralai, PhysRevB_heikkila, PhysRevB_burkov}. 
The nodal line can be associated with the non-trivial Berry phase and the semimetal supports unusual surface states, for a recent review \cite{nodal_line1}. The existence of such band structure may be protected by time reversal, 
spatial inversion
or mirror symmetries \cite{PhysRevLett.115.036806, PhysRevB.92.081201,PhysRevB.93.205132}. A variety of materials have been reported \cite{bian_TlTaSe2_PRB,PhysRevLett.117.016602,PhysRevB.94.121108,PhysRevB.96.161112,PhysRevX.8.031044, takane2018CaAgAsARPS,PhysRevB.96.245101}. The presence of almost non-dispersive Landau level was revealed in Ref.~\cite{PhysRevB.92.045126}. ARPES and quantum oscillation measurements have revealed the signatures of the nodal ring, see, for example \cite{PhysRevLett.117.016602,PhysRevB.95.121109} and review  \cite{RevModPhys_Experimental_nodal_ring}.

The collective excitations are the fundamental optical properties of the fermionic systems. It was noted that the features of plasmon modes in gapless Dirac material like graphene\cite{PhysRevB.75.205418} drastically differs from the electronic systems described by quadratic band structure. A similar investigation was also carried out in silicene\cite{PhysRevB.90.035142}, a two-dimensional puckered spin-orbit coupled honeycomb lattice of silicon atoms, exploring the spin and valley dependent plasmon modes. Recently, much attention has been paid to Dirac-Weyl materials in investigating plasmon modes with a particular focus on the chiral anomaly \cite{PhysRevLett.102.206412,PhysRevB.91.035114,PhysRevB.89.245103}.  Very recently, such collective modes have been studied in NLSMs as well \cite{Yong_Baek_Kim, PhysRevB.93.085138}. It was shown that the plasmon modes in three dimensional Weyl and NLSM are gapped in the long wave-length limit. 
The gap is determined by the respective density of states, which in the former case may be tuned by the parallel electric and magnetic field thanks to the chiral anomaly.  
Although, the impact of electron-electron exchange interaction in the presence of population imbalance (either valley or spin, etc.) on collective excitations in Dirac materials is far from being understood.

The importance of exchange interaction in the situation with non-equilibrium distribution of electron spin on spin-wave collective modes was first emphasized in Ref. \cite{aronov1977spin}. Later the spin-waves in metals and semiconductors have been extensively investigated theoretically in Refs.~\onlinecite{aronov1977spin,deutsch2010spin,PhysRevLett.97.047204,Zyuzin_2010,bashkin_1986}. 
Several experimental works were reported on the detection of spin-wave in gaseous spin polarized hydrogen, in the mixture of $^3He$ and $^4He$, polarized $^3He$ (see Ref. \onlinecite{bashkin_1986}). 
Recently, with analogy to spin, the exchange interaction induced valley wave in the Dirac materials with valley population imbalance has been discussed in Ref. \cite{PhysRevB.100.121402}. 
 Apart from graphene \cite{ju2011graphene,di2013observation} and Weyl semimetal \cite{PhysRevB.99.121401}, the NLSM plasmon modes have been recently observed experimentally\cite{PhysRevLett.127.186802}.
    
In this work, we consider a NLSM hosting two nodal rings with opposite spin orientation in the presence of the spin imbalance. The model of a semimetal with concentric loops that come from different spin channels in its band structure was introduced in \cite{PhysRevLett_NR_spin}. We consider transverse spin waves mediated by the electron-electron exchange interaction. In this model, we find three gapped modes and a gapless mode. These modes are anisotropic and disperse quadratically in all direction in the long-wave limit. We also obtain a gapless spin-wave mode corresponding to the surface states.\\
{\em Collective modes}
We start with the effective Hamiltonian describing a model of semimetal with two nodal-rings that have opposite spin polarization, \cite{PhysRevLett_NR_spin}
\begin{equation}\label{hamil}
H_{\eta}(\mathbf{k})=\epsilon\left(\frac{k_\perp^2}{Q^2}-1\right)\sigma^x+ v_z k_z\sigma^z
-\eta\frac{\lambda n}{2},
\end{equation}
where $\epsilon>0$ determines the top and bottom edges of the conductance and valence bands, respectively, and serves as an energy cut-off for the two-band model, $Q$ defines the radius of the nodal ring in the plane $k_z=0$, $v_z$ is the Fermi velocity along the $z$-direction, and $\sigma^{i},~ i=(x,y,z)$ are the Pauli matrices acting on the orbital space.
The last term describes the antisymmetric part of the electron-electron exchange interaction energy between two nodal rings, 
where the nodal ring spin-index is denoted by $\eta=\pm 1$. 
Note the exchange interaction is taken to be momentum independent within the simplest approximation for the Fourier component of screened Coulomb potential by $\lambda > 0$.
 
The low energy band dispersion is given by
\begin{equation}
 E_{b,\eta}=b\sqrt{\epsilon^2(k_\perp^2/Q^2 - 1)^2+v_z^2k_z^2}-\eta\frac{\lambda n}{2}
\end{equation}
where $b=\pm$ is the band index. The exchange interaction induced term splits the position of nodal rings in energy by $|\lambda n|$ with respect to each other. This is similar to the case of Dirac semimetals \cite{PhysRevB.100.121402}.

The exchange energy in Eq.~(\ref{hamil}) can be evaluated self consistently from the difference of particle densities between two nodal-rings as
\begin{equation}\label{density}
 n= \sum_{b=\pm}\int\frac{d^3k}{(2\pi)^3}\{f[E_{b,+}]-f[E_{b,-}]\},
\end{equation}
where $f[E_{b,\eta}]=[1+e^{(E_{b,\eta}-\mu_{\eta})/T}]^{-1}$ is the Fermi-Dirac distribution function in the nodal ring $\eta$ at temperature $T$. Summation is performed over two bands.
The spin imbalance is described by the spin-dependent chemical potential as 
$\mu_{\pm}=\mu\pm \delta\mu/2$. As noted above, $\epsilon > \mu$ is the largest energy scale in the problem, 
so that two-band approximation in formula (\ref{hamil}) may be valid. {For our study, we keep zero temperature $T=0$ throughout  the discussion. }

 
At $0< \mu_{\pm}\pm \lambda n/2 < \epsilon$, the particle density imbalance is linearly proportional to $\delta\mu$:
\begin{equation}\label{SCE1}
n = \frac{\nu \delta\mu}{1-\lambda\nu},
\end{equation}
where $\nu=\mu Q^2/4\pi \epsilon v_z$ is the density of states at the Fermi level.

To proceed, let us consider a transverse magnetic field $\propto e^{i(\omega t-{\bf q\cdot r})}$, which provides the electron transition between the two nodal rings with spin imbalance.
The dispersion of collective modes is given by the poles of transverse susceptibility. The retarded and advanced Green functions are given by $G_{\eta}^{R/A}({\bf k}, \omega) = [\omega -  H_{\eta}(\mathbf{k}) \pm i\delta]^{-1}$
and that can be used to evaluate the Keldysh Green function as $G_{\eta}^{K}({\bf k},\omega)=[1- 2 f_{\eta}(\omega)][G_{\eta}^{R}({\bf k},\omega)-G^{A}_{\eta}({\bf k},\omega)]$. 
To obtain the dispersion relation of the inter-ring collective modes, we follow the standard random phase approximation
and seek the poles of the transverse susceptibility as $\mathrm{det}( 1-\lambda \Pi ) = 0$, where the inter-ring polarization operator is a $4\times4$ matrix, which is given by
\begin{eqnarray}
 \Pi_{ab,cd}&=& \frac{i}{2} \int\frac{d^3k}{(2\pi)^3}\frac{d\Omega}{2\pi}\Big[G_{-,ab}^{R}(k+q,\omega+\Omega) G_{+,cd}^{K}(k,\Omega)\nonumber\\&+&G_{-,ab}^{K}(k+q,\omega+\Omega) G_{+,cd}^{A}(k,\Omega)   
        \Big].
\end{eqnarray}
It is convenient to write the polarization operator in new pseudo-spin representation $\Pi^{\alpha\beta}= \mathrm{Tr}_2\sigma_{bc}^{\alpha}\Pi_{ab,cd}\sigma_{da}^{\beta}/2$ 
where indices $\alpha,\beta$ take the values $(0,x,y,z)$, i.e, one singlet and three triplets, and $\mathrm{Tr}_2$ denotes trace of Pauli matrices. 

The polarization operator in new representation can be written through the sum of intra-band and inter-band terms as $\Pi^{\alpha\beta}=\Pi^{\alpha\beta}_{intra}+\Pi^{\alpha\beta}_{inter}$. 
The diagonal components which correspond to the inter-band contributions are obtained as
$\Pi^{00}_{inter}=0$, $\Pi^{xx}_{inter}=2\Pi^{yy}_{inter}=\Pi^{zz}_{inter}=\nu \ln(2\epsilon/\mu)/2$. On the other hand, the intra-band components are obtained as
\begin{align}
&\Pi^{00}_{intra} = -\nu\frac{\delta\mu+\lambda n}{\omega-\lambda n}-\nu\frac{\delta\mu+\omega}{(\omega-\lambda n)^3}\left(v_\perp^2q_\perp^2+\frac{v_z^2q_z^2}{2}\right),\nonumber\\
&\Pi^{xx}_{intra} = -\frac{\nu}{2}\frac{\delta\mu+\lambda n}{\omega-\lambda n}-\nu\frac{\delta\mu+\omega}{(\omega-\lambda n)^3}\left(v_\perp^2q_\perp^2+\frac{v_z^2q_z^2}{2}\right),
\end{align}
$\Pi_{intra}^{yy}=0$, and $\Pi_{intra}^{zz}=\Pi_{intra}^{xx}$, where $v_{\perp}= |\epsilon|/Q$. The off diagonal intra-band components are obtained up to linear in $q$ order as $\Pi^{0x}_{intra}=\Pi^{0y}_{intra}=0$,
$
\Pi^{0 z}_{intra}=\Pi_{intra}^{xz}=\Pi_{intra}^{yz}=\Pi_{intra}^{zz}=\nu v_zq_z/(\omega-\lambda n)
$, and $\Pi^{xy}_{intra}=0$. The off-diagonal components describe coupling between different modes.

 \begin{figure}[t]
 { \includegraphics[width=.45\textwidth,height=6.5cm]{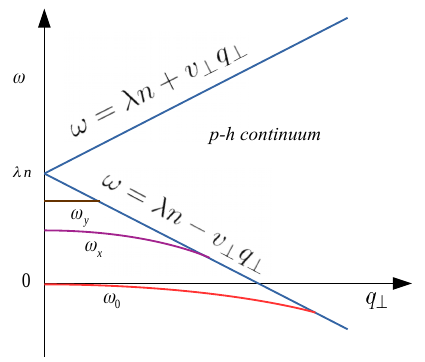}}
\caption{Dispersions of the collective modes shown for the case $q_z=0$. The particle-hole continuum is depicted by the region confined by two lines $\omega=\lambda n\pm v_\perp q_\perp$.}
\label{modes}
\end{figure}

Using the solution of the self-consistency equation (\ref{SCE1}) and assuming $\lambda n > v_{\perp} q_{\perp}, v_{z} q_{z}$ and $\lambda n > |\omega|$, the dispersion of collective modes can be obtained as
\begin{equation}\label{answer}
 \omega_0 = - \frac{1-\lambda\nu}{\lambda n} \left( v_\perp^2q_\perp^2+ \frac{1+\gamma}{2} v_z^2q_z^2 \right),
\end{equation}
where $\gamma = \lambda^2\nu^2[(1-\lambda\nu)(1-C)]^{-1} \ll 1$  with $C= 2\lambda\nu \ln(2\epsilon/\mu)$.
The other modes are given by
\begin{align}\nonumber      
&\omega_x = \frac{\lambda n}{2}\left(1-C\right)- \frac{1-\lambda\nu}{\lambda n} \left(v_\perp^2q_\perp^2+ \frac{1+\gamma}{2} v_z^2q_z^2 \right),\\
&\omega_y = \lambda n\left(1- \frac{C}{2}\right)- 2 \frac{(\lambda \nu)^2}{\lambda n}\frac{v_z^2q_z^2}{1-C}
\end{align} 
and $\omega_z=\omega_x$. We note that three modes ($\omega_x,\omega_y,\omega_z$), corresponding to the triplet components of the polarization operator are gapped whereas the singlet component ($\omega_0$) is gapless. The degree of gap is determined by the strength of electron-electron exchange interaction. 
The dispersions of collective modes are schematically shown in the Fig.~(\ref{modes}). All these modes become damped while crossing the boundary of particle-hole continuum region, which, for example for $q_z=0$, is defined by $\lambda n-v_\perp q_\perp\le\omega\le\lambda n+v_\perp q_\perp$.

Let us quickly comment on the zero pumping case when $\delta\mu=0$. First at $n = 0$,  the components of polarization operator reduce to
\begin{eqnarray}
 \Pi^{00}_{intra}&=&-\frac{\nu}{\omega^2}\left( v_\perp^2q_\perp^2+\frac{v_z^2q_z^2}{2} \right)
\end{eqnarray}
and $\Pi^{xx}_{intra}=\Pi^{yy}_{intra}=\Pi^{00}_{intra}$. The collective mode with dispersion
\begin{equation}
\omega = \sqrt{- \lambda \nu} \left( v_\perp^2q_\perp^2+\frac{v_z^2q_z^2}{2} \right)^{1/2}
\end{equation}
exists provided $\lambda < 0$.
Indeed, without the particle imbalance, the plasmon modes are expected. As it was shown in Ref.~\cite{PhysRevB.93.085138}. In this work, contrary to our study, the interaction term was taken to be 
$\lambda \sim1/q^2$ which leads to the gapped spectrum of the plasmon modes.
 
{\em Surface states}
One of the unique aspects of nodal-ring semimetals is the existence of the so-called drum-head surface states. 
Let us consider NLSM occupying a region $z>0$ with a boundary at $z=0$. 
We seek for a wave function at the boundary as $\Psi=[A, B]^{T} e^{-z\alpha + i\mathbf{k}_{\perp}\cdot \mathbf{r}}$, where $\mathrm{Re}\alpha >0$ and $A, B$ are some coefficients. The eigenvalue equation reads
\begin{equation}
[ \epsilon(k_\perp^2/Q^2 -1) \sigma^x+ i\alpha v_z \sigma^z ]\Psi= (E_{\eta}+\eta \frac{\lambda_s n_s}{2} )\Psi.
\end{equation}
Here the exchange energy at the surface is denoted by $\lambda_s$ and $E_{\eta}$ is the energy eigenvalue for nodal ring $\eta =\pm 1$. Using the hermiticity of Hamiltonian one determines the boundary condition and obtains a wave function up to an arbitrary phase $\psi$ in the form
\begin{equation}
\Psi \propto [1, e^{i\psi}]^{T} e^{-z\alpha + i\mathbf{k}_{\perp}\cdot \mathbf{r}}.
\end{equation}
Generally $\psi$ may be a function of position at the boundary. Here we consider it to be coordinate independent.
Separating the real and imaginary part from equation $[iv_z\alpha-(E_{\eta}+\eta\lambda_sn/2)] + \epsilon(k_\perp^2/Q^2 -1)e^{i\psi}=0$, one gets the dispersion of surface state in the form
\begin{equation}
 E_{\eta} = \epsilon(k_\perp^2/Q^2 -1)\cos\psi - \eta\frac{\lambda_s n_s}{2}
\end{equation}
provided condition
\begin{equation}
\alpha v_z \equiv -\epsilon(k_\perp^2/Q^2-1)\sin\psi>0
\end{equation}
holds. At $\psi = \pm \pi/2$, the flat band surface state is strictly localized at the surface where as the state is delocalized for $\psi=0$. Without loosing the generality, we consider $\psi \in [0,\pi/2)$.

Let us now discuss the collective modes for surface states in the presence of population imbalance. We neglect the interference between bulk and localized surface states, as we are looking at two limiting cases only.
Following the same approach as in the bulk we shall obtain the density imbalance at the surface. The chemical potential for the surface states with spin imbalance is taken to be $\mu_\eta=\mu+\eta\delta\mu/2$. One obtains
$
n_s = \nu_s \delta\mu/(1-\lambda_s\nu_s)
$ 
provided $\mu<0$ and $|\mu| < \epsilon\cos\psi$. Here $\nu_s=Q^2/(4\pi \epsilon \cos\psi)$ is the 2D surface density of states.  The polarization operator is obtained as
\begin{eqnarray}\label{pola_sur} 
\Pi_{s} &=& - \nu_s\frac{\delta\mu+\lambda_s n_s}{\omega- \lambda_s n_s} \\\nonumber
&-& \nu_s \Big(1 - \frac{|\mu|}{\epsilon \cos\psi}\Big)\frac{\omega + \delta\mu}{(\omega - \lambda_s n_s)^3}\left(v_\perp q_\perp \cos\psi \right)^2.
\end{eqnarray}
Hence, the surface mode dispersion at $\lambda_s\nu_s > v_\perp q_\perp$ and $\lambda_s\nu_s > |\omega|$ is given by 
\begin{equation}
 \omega=- \left(1 - \frac{|\mu|}{\epsilon \cos\psi} \right)(1- \lambda_s\nu_s) \cos^2(\psi) \frac{v_\perp^2 q_\perp^2}{\lambda_s n_s},
\end{equation}
which is quadratic in wave-vector and the slope is strongly determined by the properties of the surface. Note that similar to $\omega_0$ in Eq.~(\ref{answer}), the surface mode varies inversely with $\lambda_s n_s$.    
However, the mode strongly depends on the properties of the boundary, vanishing in the limit $\cos\psi \rightarrow  |\mu|/\epsilon$.
 Now we comment on the dispersion in absence of imbalance i.e., $\delta\mu=0$. In this case, the polarization operator reduces as
\begin{eqnarray}\label{pola_sur}
\Pi_{s} &=& -\nu_s \Big(1 - \frac{|\mu|}{\epsilon \cos\psi}\Big)\left(\frac{v_\perp q_\perp \cos\psi }{\omega}\right)^2.
\end{eqnarray}
which leads the spectrum as  
\begin{equation}
 \omega=\sqrt{-\lambda_s\nu_s}\Big(1 - \frac{|\mu|}{\epsilon \cos\psi}\Big)^{1/2}v_\perp q_\perp \cos\psi 
\end{equation}
indicating that surface mode exists only for $\lambda_s<0$. 
It is also interesting to note that the mode is linearly dispersive, whereas in presence of imbalance it disperses quadratically.\\
{\em Effects of tilt}
Now we comment on the possible effects of time-reversal symmetry breaking perturbation on the spectrum of the mode. The Hamiltonian can be written for tilted nodal ring as
\begin{equation}
H_{\eta} = \epsilon\left(\frac{k_\perp^2}{Q^2}-1\right)(\sigma^x +\eta t_{\perp})+ v_z k_z(\sigma^z+\eta t_{z}) - \eta\frac{\lambda n}{2},
\end{equation}
The spin index ($\eta$) indicates that the two nodal rings are tilted in opposite direction.
We note that for the weak tilting $|t_{\perp}| \ll 1, |t_{z}| \ll 1$, the intra-band component of the polarization operator is slightly affected in terms of the coefficients at $q_\perp, q_z$. Hence, the tilting modifies velocities $v_{\perp}$ and $v_z$ of valley-wave modes but not the wave-vector power law dependence. It is also noteworthy to mention that the weak tilting does not affect the mode gap.

Finally we draw a comparison with the valley-wave in gapless Dirac-Weyl systems, studied recently  \cite{PhysRevB.100.121402}. We note that the gapless collective mode in the presence of the population imbalance in NLSM, graphene, or Weyl semimetals exhibit almost similar behaviour except the degree of curvature, which depends on system parameters. 
However, the degree of gap in the spectrum is strongly determined by the dimension and exchange interaction. For example, we note that the gapless mode exists in the NLSM case for any sign of the relative Berry flux of the two nodal-rings.
This is not the case for three dimensional Dirac-Weyl materials as discussed in  \cite{PhysRevB.100.121402}. Another important difference with Dirac-Weyl systems is the feature of collective modes which arise due to surface states. 

Although, collective charge dynamics of Fermi arc surface states in Weyl semimetals has been studied \cite{PhysRevB_surfacestates_Sukhachov, PhysRevB_surfacestates_plasmon},   
to the best of our knowledge, the exchange interaction mediated collective modes in Fermi arcs surface states have not been discussed yet. 
To support the population imbalance scenario and resulting collective modes, the system has to posses at least two orthogonal surface states.
This criteria may be satisfied, for example, in antiferromagnetic Dirac semimetals, where counter-propagating Fermi arc surface states are expected \cite{Dirac_antiferromagnet}. 

{\em Summary}
To conclude, we investigate the electron-electron exchange interaction generated collective modes in nodal-line semimetal, which arise due to the population imbalance between two nodal rings.
We evaluate the transverse polarization function, which enables to obtain the dispersion relations of inter-node collective modes. We find a gapless mode with quadratic dispersion relation for both the bulk and surface states.  Very recently, plasmon modes have been observed in NLSM\cite{PhysRevLett.127.186802} in equilibrium situation, which can be extended further to spin-non-equilibrium cases with double/multiple nodal rings to realize our findings. We also reveal that the plasmon dispersion for surface states switches to linearly dispersive from quadratic one while turning off the imbalance between two rings.

{\it Acknowledgements:}
This work is supported by the Academy of Finland Grant No. 308339. 
  
\bibliography{Bib_valley_v1}

\begin{thebibliography}{38}%
\makeatletter
\providecommand \@ifxundefined [1]{%
 \@ifx{#1\undefined}
}%
\providecommand \@ifnum [1]{%
 \ifnum #1\expandafter \@firstoftwo
 \else \expandafter \@secondoftwo
 \fi
}%
\providecommand \@ifx [1]{%
 \ifx #1\expandafter \@firstoftwo
 \else \expandafter \@secondoftwo
 \fi
}%
\providecommand \natexlab [1]{#1}%
\providecommand \enquote  [1]{``#1''}%
\providecommand \bibnamefont  [1]{#1}%
\providecommand \bibfnamefont [1]{#1}%
\providecommand \citenamefont [1]{#1}%
\providecommand \href@noop [0]{\@secondoftwo}%
\providecommand \href [0]{\begingroup \@sanitize@url \@href}%
\providecommand \@href[1]{\@@startlink{#1}\@@href}%
\providecommand \@@href[1]{\endgroup#1\@@endlink}%
\providecommand \@sanitize@url [0]{\catcode `\\12\catcode `\$12\catcode
  `\&12\catcode `\#12\catcode `\^12\catcode `\_12\catcode `\%12\relax}%
\providecommand \@@startlink[1]{}%
\providecommand \@@endlink[0]{}%
\providecommand \url  [0]{\begingroup\@sanitize@url \@url }%
\providecommand \@url [1]{\endgroup\@href {#1}{\urlprefix }}%
\providecommand \urlprefix  [0]{URL }%
\providecommand \Eprint [0]{\href }%
\providecommand \doibase [0]{http://dx.doi.org/}%
\providecommand \selectlanguage [0]{\@gobble}%
\providecommand \bibinfo  [0]{\@secondoftwo}%
\providecommand \bibfield  [0]{\@secondoftwo}%
\providecommand \translation [1]{[#1]}%
\providecommand \BibitemOpen [0]{}%
\providecommand \bibitemStop [0]{}%
\providecommand \bibitemNoStop [0]{.\EOS\space}%
\providecommand \EOS [0]{\spacefactor3000\relax}%
\providecommand \BibitemShut  [1]{\csname bibitem#1\endcsname}%
\let\auto@bib@innerbib\@empty
\bibitem [{\citenamefont {Mikitik}\ and\ \citenamefont
  {Sharlai}(1999)}]{PhysRevLett_shralai}%
  \BibitemOpen
  \bibfield  {author} {\bibinfo {author} {\bibfnamefont {G.~P.}\ \bibnamefont
  {Mikitik}}\ and\ \bibinfo {author} {\bibfnamefont {Y.~V.}\ \bibnamefont
  {Sharlai}},\ }\href {\doibase 10.1103/PhysRevLett.82.2147} {\bibfield
  {journal} {\bibinfo  {journal} {Phys. Rev. Lett.}\ }\textbf {\bibinfo
  {volume} {82}},\ \bibinfo {pages} {2147} (\bibinfo {year}
  {1999})}\BibitemShut {NoStop}%
\bibitem [{\citenamefont {Heikkila}\ and\ \citenamefont
  {Volovik}(2011)}]{PhysRevB_heikkila}%
  \BibitemOpen
  \bibfield  {author} {\bibinfo {author} {\bibfnamefont {T.}~\bibnamefont
  {Heikkila}}\ and\ \bibinfo {author} {\bibfnamefont {G.~E.}\ \bibnamefont
  {Volovik}},\ }\href {\doibase 10.1134/S002136401102007X} {\bibfield
  {journal} {\bibinfo  {journal} {JETP Lett.}\ }\textbf {\bibinfo {volume}
  {93}},\ \bibinfo {pages} {59} (\bibinfo {year} {2011})}\BibitemShut {NoStop}%
\bibitem [{\citenamefont {Burkov}\ \emph {et~al.}(2011)\citenamefont {Burkov},
  \citenamefont {Hook},\ and\ \citenamefont {Balents}}]{PhysRevB_burkov}%
  \BibitemOpen
  \bibfield  {author} {\bibinfo {author} {\bibfnamefont {A.~A.}\ \bibnamefont
  {Burkov}}, \bibinfo {author} {\bibfnamefont {M.~D.}\ \bibnamefont {Hook}}, \
  and\ \bibinfo {author} {\bibfnamefont {L.}~\bibnamefont {Balents}},\ }\href
  {\doibase 10.1103/PhysRevB.84.235126} {\bibfield  {journal} {\bibinfo
  {journal} {Phys. Rev. B}\ }\textbf {\bibinfo {volume} {84}},\ \bibinfo
  {pages} {235126} (\bibinfo {year} {2011})}\BibitemShut {NoStop}%
\bibitem [{\citenamefont {Yang}\ \emph {et~al.}(2018)\citenamefont {Yang},
  \citenamefont {Yang}, \citenamefont {Derunova}, \citenamefont {Parkin},
  \citenamefont {Yan},\ and\ \citenamefont {Ali}}]{nodal_line1}%
  \BibitemOpen
  \bibfield  {author} {\bibinfo {author} {\bibfnamefont {S.-Y.}\ \bibnamefont
  {Yang}}, \bibinfo {author} {\bibfnamefont {H.}~\bibnamefont {Yang}}, \bibinfo
  {author} {\bibfnamefont {E.}~\bibnamefont {Derunova}}, \bibinfo {author}
  {\bibfnamefont {S.~S.~P.}\ \bibnamefont {Parkin}}, \bibinfo {author}
  {\bibfnamefont {B.}~\bibnamefont {Yan}}, \ and\ \bibinfo {author}
  {\bibfnamefont {M.~N.}\ \bibnamefont {Ali}},\ }\href {\doibase
  10.1080/23746149.2017.1414631} {\bibfield  {journal} {\bibinfo  {journal}
  {Adv. Phys.: X}\ }\textbf {\bibinfo {volume} {3}},\ \bibinfo {pages}
  {1414631} (\bibinfo {year} {2018})}\BibitemShut {NoStop}%
\bibitem [{\citenamefont {Kim}\ \emph {et~al.}(2015)\citenamefont {Kim},
  \citenamefont {Wieder}, \citenamefont {Kane},\ and\ \citenamefont
  {Rappe}}]{PhysRevLett.115.036806}%
  \BibitemOpen
  \bibfield  {author} {\bibinfo {author} {\bibfnamefont {Y.}~\bibnamefont
  {Kim}}, \bibinfo {author} {\bibfnamefont {B.~J.}\ \bibnamefont {Wieder}},
  \bibinfo {author} {\bibfnamefont {C.~L.}\ \bibnamefont {Kane}}, \ and\
  \bibinfo {author} {\bibfnamefont {A.~M.}\ \bibnamefont {Rappe}},\ }\href
  {\doibase 10.1103/PhysRevLett.115.036806} {\bibfield  {journal} {\bibinfo
  {journal} {Phys. Rev. Lett.}\ }\textbf {\bibinfo {volume} {115}},\ \bibinfo
  {pages} {036806} (\bibinfo {year} {2015})}\BibitemShut {NoStop}%
\bibitem [{\citenamefont {Fang}\ \emph {et~al.}(2015)\citenamefont {Fang},
  \citenamefont {Chen}, \citenamefont {Kee},\ and\ \citenamefont
  {Fu}}]{PhysRevB.92.081201}%
  \BibitemOpen
  \bibfield  {author} {\bibinfo {author} {\bibfnamefont {C.}~\bibnamefont
  {Fang}}, \bibinfo {author} {\bibfnamefont {Y.}~\bibnamefont {Chen}}, \bibinfo
  {author} {\bibfnamefont {H.-Y.}\ \bibnamefont {Kee}}, \ and\ \bibinfo
  {author} {\bibfnamefont {L.}~\bibnamefont {Fu}},\ }\href {\doibase
  10.1103/PhysRevB.92.081201} {\bibfield  {journal} {\bibinfo  {journal} {Phys.
  Rev. B}\ }\textbf {\bibinfo {volume} {92}},\ \bibinfo {pages} {081201}
  (\bibinfo {year} {2015})}\BibitemShut {NoStop}%
\bibitem [{\citenamefont {Chan}\ \emph {et~al.}(2016)\citenamefont {Chan},
  \citenamefont {Chiu}, \citenamefont {Chou},\ and\ \citenamefont
  {Schnyder}}]{PhysRevB.93.205132}%
  \BibitemOpen
  \bibfield  {author} {\bibinfo {author} {\bibfnamefont {Y.-H.}\ \bibnamefont
  {Chan}}, \bibinfo {author} {\bibfnamefont {C.-K.}\ \bibnamefont {Chiu}},
  \bibinfo {author} {\bibfnamefont {M.~Y.}\ \bibnamefont {Chou}}, \ and\
  \bibinfo {author} {\bibfnamefont {A.~P.}\ \bibnamefont {Schnyder}},\ }\href
  {\doibase 10.1103/PhysRevB.93.205132} {\bibfield  {journal} {\bibinfo
  {journal} {Phys. Rev. B}\ }\textbf {\bibinfo {volume} {93}},\ \bibinfo
  {pages} {205132} (\bibinfo {year} {2016})}\BibitemShut {NoStop}%
\bibitem [{\citenamefont {Bian}\ \emph {et~al.}(2016)\citenamefont {Bian},
  \citenamefont {Chang}, \citenamefont {Zheng}, \citenamefont {Velury},
  \citenamefont {Xu}, \citenamefont {Neupert}, \citenamefont {Chiu},
  \citenamefont {Huang}, \citenamefont {Sanchez}, \citenamefont {Belopolski},
  \citenamefont {Alidoust}, \citenamefont {Chen}, \citenamefont {Chang},
  \citenamefont {Bansil}, \citenamefont {Jeng}, \citenamefont {Lin},\ and\
  \citenamefont {Hasan}}]{bian_TlTaSe2_PRB}%
  \BibitemOpen
  \bibfield  {author} {\bibinfo {author} {\bibfnamefont {G.}~\bibnamefont
  {Bian}}, \bibinfo {author} {\bibfnamefont {T.-R.}\ \bibnamefont {Chang}},
  \bibinfo {author} {\bibfnamefont {H.}~\bibnamefont {Zheng}}, \bibinfo
  {author} {\bibfnamefont {S.}~\bibnamefont {Velury}}, \bibinfo {author}
  {\bibfnamefont {S.-Y.}\ \bibnamefont {Xu}}, \bibinfo {author} {\bibfnamefont
  {T.}~\bibnamefont {Neupert}}, \bibinfo {author} {\bibfnamefont {C.-K.}\
  \bibnamefont {Chiu}}, \bibinfo {author} {\bibfnamefont {S.-M.}\ \bibnamefont
  {Huang}}, \bibinfo {author} {\bibfnamefont {D.~S.}\ \bibnamefont {Sanchez}},
  \bibinfo {author} {\bibfnamefont {I.}~\bibnamefont {Belopolski}}, \bibinfo
  {author} {\bibfnamefont {N.}~\bibnamefont {Alidoust}}, \bibinfo {author}
  {\bibfnamefont {P.-J.}\ \bibnamefont {Chen}}, \bibinfo {author}
  {\bibfnamefont {G.}~\bibnamefont {Chang}}, \bibinfo {author} {\bibfnamefont
  {A.}~\bibnamefont {Bansil}}, \bibinfo {author} {\bibfnamefont {H.-T.}\
  \bibnamefont {Jeng}}, \bibinfo {author} {\bibfnamefont {H.}~\bibnamefont
  {Lin}}, \ and\ \bibinfo {author} {\bibfnamefont {M.~Z.}\ \bibnamefont
  {Hasan}},\ }\href {\doibase 10.1103/PhysRevB.93.121113} {\bibfield  {journal}
  {\bibinfo  {journal} {Phys. Rev. B}\ }\textbf {\bibinfo {volume} {93}},\
  \bibinfo {pages} {121113} (\bibinfo {year} {2016})}\BibitemShut {NoStop}%
\bibitem [{\citenamefont {Hu}\ \emph {et~al.}(2016)\citenamefont {Hu},
  \citenamefont {Tang}, \citenamefont {Liu}, \citenamefont {Liu}, \citenamefont
  {Zhu}, \citenamefont {Graf}, \citenamefont {Myhro}, \citenamefont {Tran},
  \citenamefont {Lau}, \citenamefont {Wei},\ and\ \citenamefont
  {Mao}}]{PhysRevLett.117.016602}%
  \BibitemOpen
  \bibfield  {author} {\bibinfo {author} {\bibfnamefont {J.}~\bibnamefont
  {Hu}}, \bibinfo {author} {\bibfnamefont {Z.}~\bibnamefont {Tang}}, \bibinfo
  {author} {\bibfnamefont {J.}~\bibnamefont {Liu}}, \bibinfo {author}
  {\bibfnamefont {X.}~\bibnamefont {Liu}}, \bibinfo {author} {\bibfnamefont
  {Y.}~\bibnamefont {Zhu}}, \bibinfo {author} {\bibfnamefont {D.}~\bibnamefont
  {Graf}}, \bibinfo {author} {\bibfnamefont {K.}~\bibnamefont {Myhro}},
  \bibinfo {author} {\bibfnamefont {S.}~\bibnamefont {Tran}}, \bibinfo {author}
  {\bibfnamefont {C.~N.}\ \bibnamefont {Lau}}, \bibinfo {author} {\bibfnamefont
  {J.}~\bibnamefont {Wei}}, \ and\ \bibinfo {author} {\bibfnamefont
  {Z.}~\bibnamefont {Mao}},\ }\href {\doibase 10.1103/PhysRevLett.117.016602}
  {\bibfield  {journal} {\bibinfo  {journal} {Phys. Rev. Lett.}\ }\textbf
  {\bibinfo {volume} {117}},\ \bibinfo {pages} {016602} (\bibinfo {year}
  {2016})}\BibitemShut {NoStop}%
\bibitem [{\citenamefont {Takane}\ \emph {et~al.}(2016)\citenamefont {Takane},
  \citenamefont {Wang}, \citenamefont {Souma}, \citenamefont {Nakayama},
  \citenamefont {Trang}, \citenamefont {Sato}, \citenamefont {Takahashi},\ and\
  \citenamefont {Ando}}]{PhysRevB.94.121108}%
  \BibitemOpen
  \bibfield  {author} {\bibinfo {author} {\bibfnamefont {D.}~\bibnamefont
  {Takane}}, \bibinfo {author} {\bibfnamefont {Z.}~\bibnamefont {Wang}},
  \bibinfo {author} {\bibfnamefont {S.}~\bibnamefont {Souma}}, \bibinfo
  {author} {\bibfnamefont {K.}~\bibnamefont {Nakayama}}, \bibinfo {author}
  {\bibfnamefont {C.~X.}\ \bibnamefont {Trang}}, \bibinfo {author}
  {\bibfnamefont {T.}~\bibnamefont {Sato}}, \bibinfo {author} {\bibfnamefont
  {T.}~\bibnamefont {Takahashi}}, \ and\ \bibinfo {author} {\bibfnamefont
  {Y.}~\bibnamefont {Ando}},\ }\href {\doibase 10.1103/PhysRevB.94.121108}
  {\bibfield  {journal} {\bibinfo  {journal} {Phys. Rev. B}\ }\textbf {\bibinfo
  {volume} {94}},\ \bibinfo {pages} {121108} (\bibinfo {year}
  {2016})}\BibitemShut {NoStop}%
\bibitem [{\citenamefont {Wang}\ \emph {et~al.}(2017)\citenamefont {Wang},
  \citenamefont {Ma}, \citenamefont {Emmanouilidou}, \citenamefont {Shen},
  \citenamefont {Hsu}, \citenamefont {Zhou}, \citenamefont {Zuo}, \citenamefont
  {Song}, \citenamefont {Xu}, \citenamefont {Wang}, \citenamefont {Huang},
  \citenamefont {Ni},\ and\ \citenamefont {Liu}}]{PhysRevB.96.161112}%
  \BibitemOpen
  \bibfield  {author} {\bibinfo {author} {\bibfnamefont {X.-B.}\ \bibnamefont
  {Wang}}, \bibinfo {author} {\bibfnamefont {X.-M.}\ \bibnamefont {Ma}},
  \bibinfo {author} {\bibfnamefont {E.}~\bibnamefont {Emmanouilidou}}, \bibinfo
  {author} {\bibfnamefont {B.}~\bibnamefont {Shen}}, \bibinfo {author}
  {\bibfnamefont {C.-H.}\ \bibnamefont {Hsu}}, \bibinfo {author} {\bibfnamefont
  {C.-S.}\ \bibnamefont {Zhou}}, \bibinfo {author} {\bibfnamefont
  {Y.}~\bibnamefont {Zuo}}, \bibinfo {author} {\bibfnamefont {R.-R.}\
  \bibnamefont {Song}}, \bibinfo {author} {\bibfnamefont {S.-Y.}\ \bibnamefont
  {Xu}}, \bibinfo {author} {\bibfnamefont {G.}~\bibnamefont {Wang}}, \bibinfo
  {author} {\bibfnamefont {L.}~\bibnamefont {Huang}}, \bibinfo {author}
  {\bibfnamefont {N.}~\bibnamefont {Ni}}, \ and\ \bibinfo {author}
  {\bibfnamefont {C.}~\bibnamefont {Liu}},\ }\href {\doibase
  10.1103/PhysRevB.96.161112} {\bibfield  {journal} {\bibinfo  {journal} {Phys.
  Rev. B}\ }\textbf {\bibinfo {volume} {96}},\ \bibinfo {pages} {161112}
  (\bibinfo {year} {2017})}\BibitemShut {NoStop}%
\bibitem [{\citenamefont {Liu}\ \emph {et~al.}(2018)\citenamefont {Liu},
  \citenamefont {Lou}, \citenamefont {Guo}, \citenamefont {Wang}, \citenamefont
  {Sun}, \citenamefont {Li}, \citenamefont {Thirupathaiah}, \citenamefont
  {Fedorov}, \citenamefont {Shen}, \citenamefont {Liu}, \citenamefont {Lei},\
  and\ \citenamefont {Wang}}]{PhysRevX.8.031044}%
  \BibitemOpen
  \bibfield  {author} {\bibinfo {author} {\bibfnamefont {Z.}~\bibnamefont
  {Liu}}, \bibinfo {author} {\bibfnamefont {R.}~\bibnamefont {Lou}}, \bibinfo
  {author} {\bibfnamefont {P.}~\bibnamefont {Guo}}, \bibinfo {author}
  {\bibfnamefont {Q.}~\bibnamefont {Wang}}, \bibinfo {author} {\bibfnamefont
  {S.}~\bibnamefont {Sun}}, \bibinfo {author} {\bibfnamefont {C.}~\bibnamefont
  {Li}}, \bibinfo {author} {\bibfnamefont {S.}~\bibnamefont {Thirupathaiah}},
  \bibinfo {author} {\bibfnamefont {A.}~\bibnamefont {Fedorov}}, \bibinfo
  {author} {\bibfnamefont {D.}~\bibnamefont {Shen}}, \bibinfo {author}
  {\bibfnamefont {K.}~\bibnamefont {Liu}}, \bibinfo {author} {\bibfnamefont
  {H.}~\bibnamefont {Lei}}, \ and\ \bibinfo {author} {\bibfnamefont
  {S.}~\bibnamefont {Wang}},\ }\href {\doibase 10.1103/PhysRevX.8.031044}
  {\bibfield  {journal} {\bibinfo  {journal} {Phys. Rev. X}\ }\textbf {\bibinfo
  {volume} {8}},\ \bibinfo {pages} {031044} (\bibinfo {year}
  {2018})}\BibitemShut {NoStop}%
\bibitem [{\citenamefont {Takane}\ \emph {et~al.}(2018)\citenamefont {Takane},
  \citenamefont {Nakayama}, \citenamefont {Souma}, \citenamefont {Wada},
  \citenamefont {Okamoto}, \citenamefont {Takenaka}, \citenamefont {Yamakawa},
  \citenamefont {Yamakage}, \citenamefont {Mitsuhashi}, \citenamefont {Horiba}
  \emph {et~al.}}]{takane2018CaAgAsARPS}%
  \BibitemOpen
  \bibfield  {author} {\bibinfo {author} {\bibfnamefont {D.}~\bibnamefont
  {Takane}}, \bibinfo {author} {\bibfnamefont {K.}~\bibnamefont {Nakayama}},
  \bibinfo {author} {\bibfnamefont {S.}~\bibnamefont {Souma}}, \bibinfo
  {author} {\bibfnamefont {T.}~\bibnamefont {Wada}}, \bibinfo {author}
  {\bibfnamefont {Y.}~\bibnamefont {Okamoto}}, \bibinfo {author} {\bibfnamefont
  {K.}~\bibnamefont {Takenaka}}, \bibinfo {author} {\bibfnamefont
  {Y.}~\bibnamefont {Yamakawa}}, \bibinfo {author} {\bibfnamefont
  {A.}~\bibnamefont {Yamakage}}, \bibinfo {author} {\bibfnamefont
  {T.}~\bibnamefont {Mitsuhashi}}, \bibinfo {author} {\bibfnamefont
  {K.}~\bibnamefont {Horiba}},  \emph {et~al.},\ }\href {\doibase
  https://doi.org/10.1038/s41535-017-0074-z} {\bibfield  {journal} {\bibinfo
  {journal} {Quantum Mater.}\ }\textbf {\bibinfo {volume} {3}},\ \bibinfo
  {pages} {1} (\bibinfo {year} {2018})}\BibitemShut {NoStop}%
\bibitem [{\citenamefont {Behrends}\ \emph {et~al.}(2017)\citenamefont
  {Behrends}, \citenamefont {Rhim}, \citenamefont {Liu}, \citenamefont
  {Grushin},\ and\ \citenamefont {Bardarson}}]{PhysRevB.96.245101}%
  \BibitemOpen
  \bibfield  {author} {\bibinfo {author} {\bibfnamefont {J.}~\bibnamefont
  {Behrends}}, \bibinfo {author} {\bibfnamefont {J.-W.}\ \bibnamefont {Rhim}},
  \bibinfo {author} {\bibfnamefont {S.}~\bibnamefont {Liu}}, \bibinfo {author}
  {\bibfnamefont {A.~G.}\ \bibnamefont {Grushin}}, \ and\ \bibinfo {author}
  {\bibfnamefont {J.~H.}\ \bibnamefont {Bardarson}},\ }\href {\doibase
  10.1103/PhysRevB.96.245101} {\bibfield  {journal} {\bibinfo  {journal} {Phys.
  Rev. B}\ }\textbf {\bibinfo {volume} {96}},\ \bibinfo {pages} {245101}
  (\bibinfo {year} {2017})}\BibitemShut {NoStop}%
\bibitem [{\citenamefont {Rhim}\ and\ \citenamefont
  {Kim}(2015)}]{PhysRevB.92.045126}%
  \BibitemOpen
  \bibfield  {author} {\bibinfo {author} {\bibfnamefont {J.-W.}\ \bibnamefont
  {Rhim}}\ and\ \bibinfo {author} {\bibfnamefont {Y.~B.}\ \bibnamefont {Kim}},\
  }\href {\doibase 10.1103/PhysRevB.92.045126} {\bibfield  {journal} {\bibinfo
  {journal} {Phys. Rev. B}\ }\textbf {\bibinfo {volume} {92}},\ \bibinfo
  {pages} {045126} (\bibinfo {year} {2015})}\BibitemShut {NoStop}%
\bibitem [{\citenamefont {Kumar}\ \emph {et~al.}(2017)\citenamefont {Kumar},
  \citenamefont {Manna}, \citenamefont {Qi}, \citenamefont {Wu}, \citenamefont
  {Wang}, \citenamefont {Yan}, \citenamefont {Felser},\ and\ \citenamefont
  {Shekhar}}]{PhysRevB.95.121109}%
  \BibitemOpen
  \bibfield  {author} {\bibinfo {author} {\bibfnamefont {N.}~\bibnamefont
  {Kumar}}, \bibinfo {author} {\bibfnamefont {K.}~\bibnamefont {Manna}},
  \bibinfo {author} {\bibfnamefont {Y.}~\bibnamefont {Qi}}, \bibinfo {author}
  {\bibfnamefont {S.-C.}\ \bibnamefont {Wu}}, \bibinfo {author} {\bibfnamefont
  {L.}~\bibnamefont {Wang}}, \bibinfo {author} {\bibfnamefont {B.}~\bibnamefont
  {Yan}}, \bibinfo {author} {\bibfnamefont {C.}~\bibnamefont {Felser}}, \ and\
  \bibinfo {author} {\bibfnamefont {C.}~\bibnamefont {Shekhar}},\ }\href
  {\doibase 10.1103/PhysRevB.95.121109} {\bibfield  {journal} {\bibinfo
  {journal} {Phys. Rev. B}\ }\textbf {\bibinfo {volume} {95}},\ \bibinfo
  {pages} {121109} (\bibinfo {year} {2017})}\BibitemShut {NoStop}%
\bibitem [{\citenamefont {Lv}\ \emph {et~al.}(2021)\citenamefont {Lv},
  \citenamefont {Qian},\ and\ \citenamefont
  {Ding}}]{RevModPhys_Experimental_nodal_ring}%
  \BibitemOpen
  \bibfield  {author} {\bibinfo {author} {\bibfnamefont {B.~Q.}\ \bibnamefont
  {Lv}}, \bibinfo {author} {\bibfnamefont {T.}~\bibnamefont {Qian}}, \ and\
  \bibinfo {author} {\bibfnamefont {H.}~\bibnamefont {Ding}},\ }\href {\doibase
  10.1103/RevModPhys.93.025002} {\bibfield  {journal} {\bibinfo  {journal}
  {Rev. Mod. Phys.}\ }\textbf {\bibinfo {volume} {93}},\ \bibinfo {pages}
  {025002} (\bibinfo {year} {2021})}\BibitemShut {NoStop}%
\bibitem [{\citenamefont {Hwang}\ and\ \citenamefont
  {Das~Sarma}(2007)}]{PhysRevB.75.205418}%
  \BibitemOpen
  \bibfield  {author} {\bibinfo {author} {\bibfnamefont {E.~H.}\ \bibnamefont
  {Hwang}}\ and\ \bibinfo {author} {\bibfnamefont {S.}~\bibnamefont
  {Das~Sarma}},\ }\href {\doibase 10.1103/PhysRevB.75.205418} {\bibfield
  {journal} {\bibinfo  {journal} {Phys. Rev. B}\ }\textbf {\bibinfo {volume}
  {75}},\ \bibinfo {pages} {205418} (\bibinfo {year} {2007})}\BibitemShut
  {NoStop}%
\bibitem [{\citenamefont {Van~Duppen}\ \emph {et~al.}(2014)\citenamefont
  {Van~Duppen}, \citenamefont {Vasilopoulos},\ and\ \citenamefont
  {Peeters}}]{PhysRevB.90.035142}%
  \BibitemOpen
  \bibfield  {author} {\bibinfo {author} {\bibfnamefont {B.}~\bibnamefont
  {Van~Duppen}}, \bibinfo {author} {\bibfnamefont {P.}~\bibnamefont
  {Vasilopoulos}}, \ and\ \bibinfo {author} {\bibfnamefont {F.~M.}\
  \bibnamefont {Peeters}},\ }\href {\doibase 10.1103/PhysRevB.90.035142}
  {\bibfield  {journal} {\bibinfo  {journal} {Phys. Rev. B}\ }\textbf {\bibinfo
  {volume} {90}},\ \bibinfo {pages} {035142} (\bibinfo {year}
  {2014})}\BibitemShut {NoStop}%
\bibitem [{\citenamefont {Das~Sarma}\ and\ \citenamefont
  {Hwang}(2009)}]{PhysRevLett.102.206412}%
  \BibitemOpen
  \bibfield  {author} {\bibinfo {author} {\bibfnamefont {S.}~\bibnamefont
  {Das~Sarma}}\ and\ \bibinfo {author} {\bibfnamefont {E.~H.}\ \bibnamefont
  {Hwang}},\ }\href {\doibase 10.1103/PhysRevLett.102.206412} {\bibfield
  {journal} {\bibinfo  {journal} {Phys. Rev. Lett.}\ }\textbf {\bibinfo
  {volume} {102}},\ \bibinfo {pages} {206412} (\bibinfo {year}
  {2009})}\BibitemShut {NoStop}%
\bibitem [{\citenamefont {Zhou}\ \emph {et~al.}(2015)\citenamefont {Zhou},
  \citenamefont {Chang},\ and\ \citenamefont {Xiao}}]{PhysRevB.91.035114}%
  \BibitemOpen
  \bibfield  {author} {\bibinfo {author} {\bibfnamefont {J.}~\bibnamefont
  {Zhou}}, \bibinfo {author} {\bibfnamefont {H.-R.}\ \bibnamefont {Chang}}, \
  and\ \bibinfo {author} {\bibfnamefont {D.}~\bibnamefont {Xiao}},\ }\href
  {\doibase 10.1103/PhysRevB.91.035114} {\bibfield  {journal} {\bibinfo
  {journal} {Phys. Rev. B}\ }\textbf {\bibinfo {volume} {91}},\ \bibinfo
  {pages} {035114} (\bibinfo {year} {2015})}\BibitemShut {NoStop}%
\bibitem [{\citenamefont {Panfilov}\ \emph {et~al.}(2014)\citenamefont
  {Panfilov}, \citenamefont {Burkov},\ and\ \citenamefont
  {Pesin}}]{PhysRevB.89.245103}%
  \BibitemOpen
  \bibfield  {author} {\bibinfo {author} {\bibfnamefont {I.}~\bibnamefont
  {Panfilov}}, \bibinfo {author} {\bibfnamefont {A.~A.}\ \bibnamefont
  {Burkov}}, \ and\ \bibinfo {author} {\bibfnamefont {D.~A.}\ \bibnamefont
  {Pesin}},\ }\href {\doibase 10.1103/PhysRevB.89.245103} {\bibfield  {journal}
  {\bibinfo  {journal} {Phys. Rev. B}\ }\textbf {\bibinfo {volume} {89}},\
  \bibinfo {pages} {245103} (\bibinfo {year} {2014})}\BibitemShut {NoStop}%
\bibitem [{\citenamefont {Rhim}\ and\ \citenamefont
  {Kim}(2016)}]{Yong_Baek_Kim}%
  \BibitemOpen
  \bibfield  {author} {\bibinfo {author} {\bibfnamefont {J.-W.}\ \bibnamefont
  {Rhim}}\ and\ \bibinfo {author} {\bibfnamefont {Y.~B.}\ \bibnamefont {Kim}},\
  }\href {\doibase 10.1088/1367-2630/18/4/043010} {\bibfield  {journal}
  {\bibinfo  {journal} {New J. Phys.}\ }\textbf {\bibinfo {volume} {18}},\
  \bibinfo {pages} {043010} (\bibinfo {year} {2016})}\BibitemShut {NoStop}%
\bibitem [{\citenamefont {Yan}\ \emph {et~al.}(2016)\citenamefont {Yan},
  \citenamefont {Huang},\ and\ \citenamefont {Wang}}]{PhysRevB.93.085138}%
  \BibitemOpen
  \bibfield  {author} {\bibinfo {author} {\bibfnamefont {Z.}~\bibnamefont
  {Yan}}, \bibinfo {author} {\bibfnamefont {P.-W.}\ \bibnamefont {Huang}}, \
  and\ \bibinfo {author} {\bibfnamefont {Z.}~\bibnamefont {Wang}},\ }\href
  {\doibase 10.1103/PhysRevB.93.085138} {\bibfield  {journal} {\bibinfo
  {journal} {Phys. Rev. B}\ }\textbf {\bibinfo {volume} {93}},\ \bibinfo
  {pages} {085138} (\bibinfo {year} {2016})}\BibitemShut {NoStop}%
\bibitem [{\citenamefont {Aronov}(1977)}]{aronov1977spin}%
  \BibitemOpen
  \bibfield  {author} {\bibinfo {author} {\bibfnamefont {A.}~\bibnamefont
  {Aronov}},\ }\href {http://www.jetp.ac.ru/cgi-bin/e/index/e/46/2/p301?a=list}
  {\bibfield  {journal} {\bibinfo  {journal} {JETP}\ }\textbf {\bibinfo
  {volume} {73}},\ \bibinfo {pages} {577} (\bibinfo {year} {1977})}\BibitemShut
  {NoStop}%
\bibitem [{\citenamefont {Deutsch}\ \emph {et~al.}(2010)\citenamefont
  {Deutsch}, \citenamefont {Ramirez-Martinez}, \citenamefont {Lacro{\^u}te},
  \citenamefont {Reinhard}, \citenamefont {Schneider}, \citenamefont {Fuchs},
  \citenamefont {Pi{\'e}chon}, \citenamefont {Lalo{\"e}}, \citenamefont
  {Reichel},\ and\ \citenamefont {Rosenbusch}}]{deutsch2010spin}%
  \BibitemOpen
  \bibfield  {author} {\bibinfo {author} {\bibfnamefont {C.}~\bibnamefont
  {Deutsch}}, \bibinfo {author} {\bibfnamefont {F.}~\bibnamefont
  {Ramirez-Martinez}}, \bibinfo {author} {\bibfnamefont {C.}~\bibnamefont
  {Lacro{\^u}te}}, \bibinfo {author} {\bibfnamefont {F.}~\bibnamefont
  {Reinhard}}, \bibinfo {author} {\bibfnamefont {T.}~\bibnamefont {Schneider}},
  \bibinfo {author} {\bibfnamefont {J.-N.}\ \bibnamefont {Fuchs}}, \bibinfo
  {author} {\bibfnamefont {F.}~\bibnamefont {Pi{\'e}chon}}, \bibinfo {author}
  {\bibfnamefont {F.}~\bibnamefont {Lalo{\"e}}}, \bibinfo {author}
  {\bibfnamefont {J.}~\bibnamefont {Reichel}}, \ and\ \bibinfo {author}
  {\bibfnamefont {P.}~\bibnamefont {Rosenbusch}},\ }\href {\doibase
  10.1103/PhysRevLett.105.020401} {\bibfield  {journal} {\bibinfo  {journal}
  {Phys. Rev. Lett.}\ }\textbf {\bibinfo {volume} {105}},\ \bibinfo {pages}
  {020401} (\bibinfo {year} {2010})}\BibitemShut {NoStop}%
\bibitem [{\citenamefont {Bedell}\ and\ \citenamefont
  {Dahal}(2006)}]{PhysRevLett.97.047204}%
  \BibitemOpen
  \bibfield  {author} {\bibinfo {author} {\bibfnamefont {K.~S.}\ \bibnamefont
  {Bedell}}\ and\ \bibinfo {author} {\bibfnamefont {H.~P.}\ \bibnamefont
  {Dahal}},\ }\href {\doibase 10.1103/PhysRevLett.97.047204} {\bibfield
  {journal} {\bibinfo  {journal} {Phys. Rev. Lett.}\ }\textbf {\bibinfo
  {volume} {97}},\ \bibinfo {pages} {047204} (\bibinfo {year}
  {2006})}\BibitemShut {NoStop}%
\bibitem [{\citenamefont {Zyuzin}\ and\ \citenamefont
  {Zyuzin}(2010)}]{Zyuzin_2010}%
  \BibitemOpen
  \bibfield  {author} {\bibinfo {author} {\bibfnamefont {A.~A.}\ \bibnamefont
  {Zyuzin}}\ and\ \bibinfo {author} {\bibfnamefont {A.~Y.}\ \bibnamefont
  {Zyuzin}},\ }\href {\doibase 10.1209/0295-5075/90/67007} {\bibfield
  {journal} {\bibinfo  {journal} {{EPL} (Europhysics Letters)}\ }\textbf
  {\bibinfo {volume} {90}},\ \bibinfo {pages} {67007} (\bibinfo {year}
  {2010})}\BibitemShut {NoStop}%
\bibitem [{\citenamefont {Bashkin}(1986)}]{bashkin_1986}%
  \BibitemOpen
  \bibfield  {author} {\bibinfo {author} {\bibfnamefont {E.}~\bibnamefont
  {Bashkin}},\ }\href@noop {} {\bibfield  {journal} {\bibinfo  {journal}
  {Soviet Physics Uspekhi}\ }\textbf {\bibinfo {volume} {29}},\ \bibinfo
  {pages} {238} (\bibinfo {year} {1986})}\BibitemShut {NoStop}%
\bibitem [{\citenamefont {Zyuzin}\ and\ \citenamefont
  {Zyuzin}(2019)}]{PhysRevB.100.121402}%
  \BibitemOpen
  \bibfield  {author} {\bibinfo {author} {\bibfnamefont {A.~A.}\ \bibnamefont
  {Zyuzin}}\ and\ \bibinfo {author} {\bibfnamefont {A.~Y.}\ \bibnamefont
  {Zyuzin}},\ }\href {\doibase 10.1103/PhysRevB.100.121402} {\bibfield
  {journal} {\bibinfo  {journal} {Phys. Rev. B}\ }\textbf {\bibinfo {volume}
  {100}},\ \bibinfo {pages} {121402} (\bibinfo {year} {2019})}\BibitemShut
  {NoStop}%
\bibitem [{\citenamefont {Ju}\ \emph {et~al.}(2011)\citenamefont {Ju},
  \citenamefont {Geng}, \citenamefont {Horng}, \citenamefont {Girit},
  \citenamefont {Martin}, \citenamefont {Hao}, \citenamefont {Bechtel},
  \citenamefont {Liang}, \citenamefont {Zettl}, \citenamefont {Shen} \emph
  {et~al.}}]{ju2011graphene}%
  \BibitemOpen
  \bibfield  {author} {\bibinfo {author} {\bibfnamefont {L.}~\bibnamefont
  {Ju}}, \bibinfo {author} {\bibfnamefont {B.}~\bibnamefont {Geng}}, \bibinfo
  {author} {\bibfnamefont {J.}~\bibnamefont {Horng}}, \bibinfo {author}
  {\bibfnamefont {C.}~\bibnamefont {Girit}}, \bibinfo {author} {\bibfnamefont
  {M.}~\bibnamefont {Martin}}, \bibinfo {author} {\bibfnamefont
  {Z.}~\bibnamefont {Hao}}, \bibinfo {author} {\bibfnamefont {H.~A.}\
  \bibnamefont {Bechtel}}, \bibinfo {author} {\bibfnamefont {X.}~\bibnamefont
  {Liang}}, \bibinfo {author} {\bibfnamefont {A.}~\bibnamefont {Zettl}},
  \bibinfo {author} {\bibfnamefont {Y.~R.}\ \bibnamefont {Shen}},  \emph
  {et~al.},\ }\href@noop {} {\bibfield  {journal} {\bibinfo  {journal} {Nature
  nanotechnology}\ }\textbf {\bibinfo {volume} {6}},\ \bibinfo {pages} {630}
  (\bibinfo {year} {2011})}\BibitemShut {NoStop}%
\bibitem [{\citenamefont {Di~Pietro}\ \emph {et~al.}(2013)\citenamefont
  {Di~Pietro}, \citenamefont {Ortolani}, \citenamefont {Limaj}, \citenamefont
  {Di~Gaspare}, \citenamefont {Giliberti}, \citenamefont {Giorgianni},
  \citenamefont {Brahlek}, \citenamefont {Bansal}, \citenamefont {Koirala},
  \citenamefont {Oh} \emph {et~al.}}]{di2013observation}%
  \BibitemOpen
  \bibfield  {author} {\bibinfo {author} {\bibfnamefont {P.}~\bibnamefont
  {Di~Pietro}}, \bibinfo {author} {\bibfnamefont {M.}~\bibnamefont {Ortolani}},
  \bibinfo {author} {\bibfnamefont {O.}~\bibnamefont {Limaj}}, \bibinfo
  {author} {\bibfnamefont {A.}~\bibnamefont {Di~Gaspare}}, \bibinfo {author}
  {\bibfnamefont {V.}~\bibnamefont {Giliberti}}, \bibinfo {author}
  {\bibfnamefont {F.}~\bibnamefont {Giorgianni}}, \bibinfo {author}
  {\bibfnamefont {M.}~\bibnamefont {Brahlek}}, \bibinfo {author} {\bibfnamefont
  {N.}~\bibnamefont {Bansal}}, \bibinfo {author} {\bibfnamefont
  {N.}~\bibnamefont {Koirala}}, \bibinfo {author} {\bibfnamefont
  {S.}~\bibnamefont {Oh}},  \emph {et~al.},\ }\href@noop {} {\bibfield
  {journal} {\bibinfo  {journal} {Nature nanotechnology}\ }\textbf {\bibinfo
  {volume} {8}},\ \bibinfo {pages} {556} (\bibinfo {year} {2013})}\BibitemShut
  {NoStop}%
\bibitem [{\citenamefont {Chiarello}\ \emph {et~al.}(2019)\citenamefont
  {Chiarello}, \citenamefont {Hofmann}, \citenamefont {Li}, \citenamefont
  {Fabio}, \citenamefont {Guo}, \citenamefont {Chen}, \citenamefont
  {Das~Sarma},\ and\ \citenamefont {Politano}}]{PhysRevB.99.121401}%
  \BibitemOpen
  \bibfield  {author} {\bibinfo {author} {\bibfnamefont {G.}~\bibnamefont
  {Chiarello}}, \bibinfo {author} {\bibfnamefont {J.}~\bibnamefont {Hofmann}},
  \bibinfo {author} {\bibfnamefont {Z.}~\bibnamefont {Li}}, \bibinfo {author}
  {\bibfnamefont {V.}~\bibnamefont {Fabio}}, \bibinfo {author} {\bibfnamefont
  {L.}~\bibnamefont {Guo}}, \bibinfo {author} {\bibfnamefont {X.}~\bibnamefont
  {Chen}}, \bibinfo {author} {\bibfnamefont {S.}~\bibnamefont {Das~Sarma}}, \
  and\ \bibinfo {author} {\bibfnamefont {A.}~\bibnamefont {Politano}},\ }\href
  {\doibase 10.1103/PhysRevB.99.121401} {\bibfield  {journal} {\bibinfo
  {journal} {Phys. Rev. B}\ }\textbf {\bibinfo {volume} {99}},\ \bibinfo
  {pages} {121401} (\bibinfo {year} {2019})}\BibitemShut {NoStop}%
\bibitem [{\citenamefont {Xue}\ \emph {et~al.}(2021)\citenamefont {Xue},
  \citenamefont {Wang}, \citenamefont {Li}, \citenamefont {Zhang},
  \citenamefont {Jia}, \citenamefont {Zhou}, \citenamefont {Shi}, \citenamefont
  {Zhu}, \citenamefont {Yao},\ and\ \citenamefont
  {Guo}}]{PhysRevLett.127.186802}%
  \BibitemOpen
  \bibfield  {author} {\bibinfo {author} {\bibfnamefont {S.}~\bibnamefont
  {Xue}}, \bibinfo {author} {\bibfnamefont {M.}~\bibnamefont {Wang}}, \bibinfo
  {author} {\bibfnamefont {Y.}~\bibnamefont {Li}}, \bibinfo {author}
  {\bibfnamefont {S.}~\bibnamefont {Zhang}}, \bibinfo {author} {\bibfnamefont
  {X.}~\bibnamefont {Jia}}, \bibinfo {author} {\bibfnamefont {J.}~\bibnamefont
  {Zhou}}, \bibinfo {author} {\bibfnamefont {Y.}~\bibnamefont {Shi}}, \bibinfo
  {author} {\bibfnamefont {X.}~\bibnamefont {Zhu}}, \bibinfo {author}
  {\bibfnamefont {Y.}~\bibnamefont {Yao}}, \ and\ \bibinfo {author}
  {\bibfnamefont {J.}~\bibnamefont {Guo}},\ }\href {\doibase
  10.1103/PhysRevLett.127.186802} {\bibfield  {journal} {\bibinfo  {journal}
  {Phys. Rev. Lett.}\ }\textbf {\bibinfo {volume} {127}},\ \bibinfo {pages}
  {186802} (\bibinfo {year} {2021})}\BibitemShut {NoStop}%
\bibitem [{\citenamefont {Zhang}\ \emph {et~al.}(2020)\citenamefont {Zhang},
  \citenamefont {Zhang}, \citenamefont {Liu},\ and\ \citenamefont
  {Yao}}]{PhysRevLett_NR_spin}%
  \BibitemOpen
  \bibfield  {author} {\bibinfo {author} {\bibfnamefont {R.-W.}\ \bibnamefont
  {Zhang}}, \bibinfo {author} {\bibfnamefont {Z.}~\bibnamefont {Zhang}},
  \bibinfo {author} {\bibfnamefont {C.-C.}\ \bibnamefont {Liu}}, \ and\
  \bibinfo {author} {\bibfnamefont {Y.}~\bibnamefont {Yao}},\ }\href {\doibase
  10.1103/PhysRevLett.124.016402} {\bibfield  {journal} {\bibinfo  {journal}
  {Phys. Rev. Lett.}\ }\textbf {\bibinfo {volume} {124}},\ \bibinfo {pages}
  {016402} (\bibinfo {year} {2020})}\BibitemShut {NoStop}%
\bibitem [{\citenamefont {Gorbar}\ \emph {et~al.}(2019)\citenamefont {Gorbar},
  \citenamefont {Miransky}, \citenamefont {Shovkovy},\ and\ \citenamefont
  {Sukhachov}}]{PhysRevB_surfacestates_Sukhachov}%
  \BibitemOpen
  \bibfield  {author} {\bibinfo {author} {\bibfnamefont {E.~V.}\ \bibnamefont
  {Gorbar}}, \bibinfo {author} {\bibfnamefont {V.~A.}\ \bibnamefont
  {Miransky}}, \bibinfo {author} {\bibfnamefont {I.~A.}\ \bibnamefont
  {Shovkovy}}, \ and\ \bibinfo {author} {\bibfnamefont {P.~O.}\ \bibnamefont
  {Sukhachov}},\ }\href {\doibase 10.1103/PhysRevB.99.155120} {\bibfield
  {journal} {\bibinfo  {journal} {Phys. Rev. B}\ }\textbf {\bibinfo {volume}
  {99}},\ \bibinfo {pages} {155120} (\bibinfo {year} {2019})}\BibitemShut
  {NoStop}%
\bibitem [{\citenamefont {Adinehvand}\ \emph {et~al.}(2019)\citenamefont
  {Adinehvand}, \citenamefont {Faraei}, \citenamefont {Farajollahpour},\ and\
  \citenamefont {Jafari}}]{PhysRevB_surfacestates_plasmon}%
  \BibitemOpen
  \bibfield  {author} {\bibinfo {author} {\bibfnamefont {F.}~\bibnamefont
  {Adinehvand}}, \bibinfo {author} {\bibfnamefont {Z.}~\bibnamefont {Faraei}},
  \bibinfo {author} {\bibfnamefont {T.}~\bibnamefont {Farajollahpour}}, \ and\
  \bibinfo {author} {\bibfnamefont {S.~A.}\ \bibnamefont {Jafari}},\ }\href
  {\doibase 10.1103/PhysRevB.100.195408} {\bibfield  {journal} {\bibinfo
  {journal} {Phys. Rev. B}\ }\textbf {\bibinfo {volume} {100}},\ \bibinfo
  {pages} {195408} (\bibinfo {year} {2019})}\BibitemShut {NoStop}%
\bibitem [{\citenamefont {Tang}\ \emph {et~al.}(2016)\citenamefont {Tang},
  \citenamefont {Zhou}, \citenamefont {Xu},\ and\ \citenamefont
  {Zhang}}]{Dirac_antiferromagnet}%
  \BibitemOpen
  \bibfield  {author} {\bibinfo {author} {\bibfnamefont {P.}~\bibnamefont
  {Tang}}, \bibinfo {author} {\bibfnamefont {Q.}~\bibnamefont {Zhou}}, \bibinfo
  {author} {\bibfnamefont {G.}~\bibnamefont {Xu}}, \ and\ \bibinfo {author}
  {\bibfnamefont {S.-C.}\ \bibnamefont {Zhang}},\ }\href {\doibase
  10.1038/nphys3839} {\bibfield  {journal} {\bibinfo  {journal} {Nature
  Physics}\ }\textbf {\bibinfo {volume} {12}},\ \bibinfo {pages} {1100}
  (\bibinfo {year} {2016})}\BibitemShut {NoStop}%
\end{thebibliography}%

 \end{document}